# Evolution of interface magnetism in Fe/Alq$_3$ bilayer

Avinash Ganesh Khanderao[1,2], Sonia Kaushik[1], Arun Singh Dev[1], V.R. Reddy[1], Ilya Sergueev[3], Hans-Christian Wille[3], Pallavi Pandit[3], Stephan V Roth[3] and Dileep Kumar[1,a]

[1] UGC-DAE Consortium for Scientific Research, Khandwa Road, Indore-452001, India
[2] Department of Physics, Jagadamba Mahavidyalaya, Achalpur City-444806, India
[3] Deutsches Elektronen-Synchrotron (DESY), Notkestrasse 85, D-22607 Hamburg, Germany
[a] Corresponding author: dkumar@csr.res.in

## ABSTRACT

This work provides a better understanding and behavior of the metal-organic interface magnetism, which plays a crucial role in organic spin-valve devices. Interface magnetism and topological structure of Fe on organic semiconductor film Tris(8-hydroxyquinolinato) aluminium (Alq$_3$) have been studied and compared with Fe film deposited directly on Si (100) substrate. To get information of the diffused Fe layer at the Fe/ Alq$_3$ interface, grazing incident nuclear resonance scattering (GINRS) measurements are made depth selective by introducing a 95% enriched thin $^{57}$Fe layer at the Interface and producing x-ray standing wave within the layered structure. Compared with Fe growth on Si substrate, where film exhibits a hyperfine field value of ~32 T (Bulk Fe), a thick Fe- Alq$_3$ interface has been found with reduced electron density and hyperfine fields providing evidence of deep penetration of Fe atoms into Alq$_3$ film. Due to the soft nature of Alq$_3$, Fe moments relax in the film plane. At the same time, Fe on Si has resultant ~43° out of plane orientation of Fe moments at the Interface due to stressed and rough Fe layer near Si. Evolution of magnetism at Fe-Alq3 Interface is monitored using in-situ magneto-optical Kerr effect (MOKE) during growth of Fe on Alq$_3$ surface and small-angle x-ray scattering (SAXS) measurements. It is found that the Fe atom tries to organize into clusters to minimize their surface/interface energy. The origin of the 24 Å thick magnetic dead layer at the Interface is attributed to the small Fe clusters of paramagnetic or superparamagnetic nature. Present work provides an understanding of interfacial magnetism at metal-organic interfaces and the topological study using GI-NRS technique, which is made depth selective to probe magnetism of the diffused ferromagnetic layer, which is otherwise difficult for lab-based techniques.

## 1. INTRODUCTION

Metal-organic interfaces play an important role in organic Spintronics based devices such as organic spin valves[1,2]. Organic Spintronics is one of the most interesting developments due to its structural flexibility, low production cost, long spin relaxation time and low spin-orbital coupling[3,4], Due to the large electron spin relaxation time, it is an emerging area for the future electronic devices. Since its discovery, significant progress has been made in understanding the spin injection, manipulation, and detection in organic spin valve (OSV)[5]. The low spin-orbit coupling and weak hyperfine interaction allow the spin polarization of the carriers to retain for very long times, allowing spins to be manipulated, which makes OSC a potential candidate for organic Spintronics[6,7]. However, these potential opportunities are accompanied by impediments such as low mobility[8], layer roughness, and intermixing at the Interface strongly contribute to spin scattering and poor spin injection efficiencies[9,10]. Also, due to the mechanical softness and complicated transport at the FM-organic semiconductor (OSC) interfaces arising from metal atom penetration and diffusion[11] or possible chemical reaction at the interfaces, a clear understanding of how effective OSV could be achieved has not been experimentally realized to date. Furthermore, penetrated metal atoms in OSC can react chemically with it and can affect its transport properties. In extreme cases, top FM layer atoms penetrate deep into the organic layer [12, 13, 14, 15], causing a conductive pathway to the bottom electrode and leading to the spin valve failure.

Considerable work on various FM/OSC/FM spin valve structures (where FM= Co, Fe, LSMO, Ni, Mn, $Fe_2O_3$, etc. OSC=$Alq_3$, rubrene, C60, pentacene-T6 etc.) are available in the literature, where significant efforts have been made to understand the spin injection, manipulation, and detection in organic spin valve (OSV), effect of barrier layer at Interface and electron structures at interfaces [3,9, 10, 16,17]. Although most of the studies pointed out the importance of the top electrode interface (FM/OSC) in terms of magnetism, results are missing, where the depth profiling of the magnetic top electrode (FM/OSC) was done properly with an aim to study interface magnetism. Due to the soft nature of the film, the main problem in this area is to determine the interface structure of FM/ OSC structure accurately and to correlate the same with magnetic properties unambiguously. A variety of lab-based techniques such as SQUID VSM, magneto-optical Kerr effect (MOKE), nuclear magnetic resonance (NMR) are available. However, most of the techniques either do not have sufficient depth resolution to resolve the interfaces or may not be probing the true interfaces. Conversion electron Mossbauer spectroscopy (CEMS)[18], polarized neutron reflectivity (PNR)[19], and X-ray magnetic circular dichroism (XMCD)[20] are more informative and powerful methods for Interface resolved studies. But each technique has its own advantages and disadvantages, which mainly depends on the type of sample and the nature of magnetic studies. For example, in the case of CEMS, depth selectivity can be achieved with a thin probe resonant layer embedded at a definite depth[21,22]. For the study of depth-dependent magnetism, a set of samples need to be prepared with a probe resonant layer at different depth positions. But, even in the best deposition methods available, full reproducibility is difficult to achieve in separate depositions. In general, PNR and resonant x-ray scattering are powerful and unique tools to study magnetism and, therefore, are extensively utilized to solve complicated magnetic structures. But for ultra-thin films (fraction of nm to several nms) regime, in general, PNR is also not a good option due to relatively very long measurement time. In addition, it usually requires a bigger sample size to have sufficient neutron scattering from the Interface. Yaohua Liu, et al.,[23] have studied $Co/Alq_3/Fe$ trilayers, they mentioned the less sensitivity of the PNR to the top Interface in $Co/Alq_3/Fe$ due to less thickness of top Co as compared to bottom Fe thickness (250 Å). Unexpectedly, they have obtained an ill-defined layer at the bottom ($Alq_3$/Fe), which is in contrast to the literature [24, 25], where an ill-defined layer is found at the top Interface. In fact, they themselves mentioned in this paper that "further study of the interfacial magnetization of these transition metal films adjacent to OSCs would be worthwhile". In the case of X-ray absorption spectroscopy (XAS) and XMCD, TEY and TFY modes are used to probe the surface and Interface of thin films. TEY mode is surface sensitive, and hence interface information can be achieved by in-situ experiments or by reducing the total thickness of the film by ion beam sputtering (IBS)[26, 27]. Unfortunately, in the case of FM/OSC layers, IBS cannot be used due to the soft nature of the organic thin film. In photon-in/photon-out mode, considered to be a probe of bulk properties, up to a depths of the order of 1000 Å, although, in fact, the penetration and escape depths of the resonant x rays can be significantly reduced for excitations at a strong absorption edge of a majority elemental constituent [28] and thus the actual sensing depth is somehow ill-defined and variable from sample to sample.

On the other hand, the grazing incident nuclear resonance (GI-NRS) technique is based on third-generation synchrotron radiation sources [ESRF[29], APS [30], SPRING-8 [31], and KEK[32]] and detects reflected resonant count from the thin sample. Besides its limitation of isotopes selectivity to the limited materials ($^{57}$Fe, $^{119}$Sn, $^{121}$Sb, $^{125}$Te, $^{193}$Ir etc.)[33], the NRS technique has its advantages of high scattering yield of NRS, which makes it possible to measure even a fraction of a monolayer of isotope material within a reasonable time[34]. It is a powerful technique to extract magnetic and morphological information of the buried structures by depositing ultra-thin $^{57}$Fe layer at FM-OSC interface, which seems to be promising in order to develop a better understanding of OSV structure. As the synchrotron beam is polarized, GI-NRS is highly sensitive to magnetization directions in-plane and out-of-plane without extra experimental setups. In the present case, Fe-Alq$_3$ interfaces are studied using the GI-NRS technique. GI-NRS contribution was enhanced from diffused part of the Fe layer under x-ray standing wave (XSW) condition. The evolution of the Fe magnetism on the Alq$_3$ surface is also studied by performing in-situ MOKE measurements under ultra-high vacuum (UHV) conditions. Combined analysis of ex-situ as well in-situ data is used to develop a better understanding of the Fe/Alq$_3$ interfaces.

## 2. EXPERIMENTAL

The thin film of organic semiconductor (Alq$_3$) was deposited using thermal evaporation (TE) in the high vacuum chamber at the base vacuum of $1\times10^{-6}$ mbar. The Alq$_3$ layers were sublimated from a commercial Alq$_3$ powder (sublimated grade, >99% pure, supplied by Sigma Aldrich) at a constant rate of 0.02 nm s$^{-1}$ using TE technique. During the deposition of Alq$_3$, the crucible temperature was monitored and did not vary by more than ±1°C. In-situ MOKE measurements were performed at the base pressure of $2\times10^{-9}$ mbar. Whereas the Fe layer is deposited on the Alq$_3$ layer using electron beam (EB) evaporation in a separate UHV chamber. Film thickness was monitored using a calibrated quartz crystal thickness monitor. The sample was aligned such that the real-time MOKE in longitudinal geometry and transport measurements were performed simultaneously[35, 36, 37,38]. MOKE components are attached with the chamber, where He-Ne laser with monochromatic wavelength 6328 Å is used for the measurements. A magnetic field of 0.15T is generated using an in-situ electromagnet. NRS is sensitive to isotopes, therefore two sets of samples a) (a) Si/$^{57}$Fe/Fe (denoted as SIF) and (b) Si/Pt/Alq$_3$/$^{57}$Fe /Fe (denoted as ALF) were also prepared by depositing the $^{57}$Fe (95% enriched) marker layer at the Interface using electron beam evaporation technique under UHV condition. This is done with an aim to increase interface selectivity for the NRS technique. In the case of the second sample, $^{57}$Fe atoms are expected to diffuse deep into the Alq$_3$ later, therefore resonant $^{57}$Fe nuclei per unit volume drastically decrease. A few angstroms thick $^{57}$Fe layer may spread over up to several nanometers in the Alq$_3$ layer (significantly less density of $^{57}$Fe resonant nuclei). Therefore, along with depth selectivity in GI-NRS measurement, the enhanced contribution from diffused part is achieved through coinciding the antinode part of XSW with diffused layer by making waveguide structure to generate XSW. For this purpose, Alq$_3$/$^{57}$Fe/Fe structure was deposited on a high dense Pt buffer layer[28]. The GI-NRS measurements were carried out at the nuclear resonance beamline at P01 beamline, PETRA III, DESY using photon beam energy of 14.41 keV to excite the $^{57}$Fe nuclei in the sample[39]. During measurements, the synchrotron was operating in the 40-bunch mode with bunch separation of 192 ns. The detector used in the experiment was an avalanche photodiode, which has a time resolution of ~ 1 ns. GISAXS and GIWAXS measurements were carried out at P03 beamline, PETRA III, at a photon energy of 13 KeV with a beam size of $48 \times 5$ (μm)$^2$ (horizontal to vertical ratio). PILATUS 300K pixel detector for GIWAXS and PILATUS 1M pixel detector (readout time < 3 ms and pixel size 172 μm) was used for GISAXS measurements.

## 3. RESULTS AND DISCUSSION

As prepared samples were characterized further, to get the information of the structure of the sample, XRR measurements were done on (a) SIF and (b) ALF samples as a function of scattering vector q$_x$ (increasing incident angle). Figure 1(a) gives XRR patterns together with the best fit to the data using Parratt's formalism[40]. The electron density depth profile of samples SIF and ALF are shown in fig. 1(b) and 1(c). The thickness of the $^{57}$Fe+Fe layers in both samples is almost the same (~120Å). It may be noted $^{57}$Fe and Fe are chemically the same (same electron density in both isotopes), therefore the XRR technique gives combined information of $^{57}$Fe and Fe layers. As $^{57}$Fe and Fe layers are deposited in identical conditions, therefore thicknesses of $^{57}$Fe are also expected to be the same in both samples.

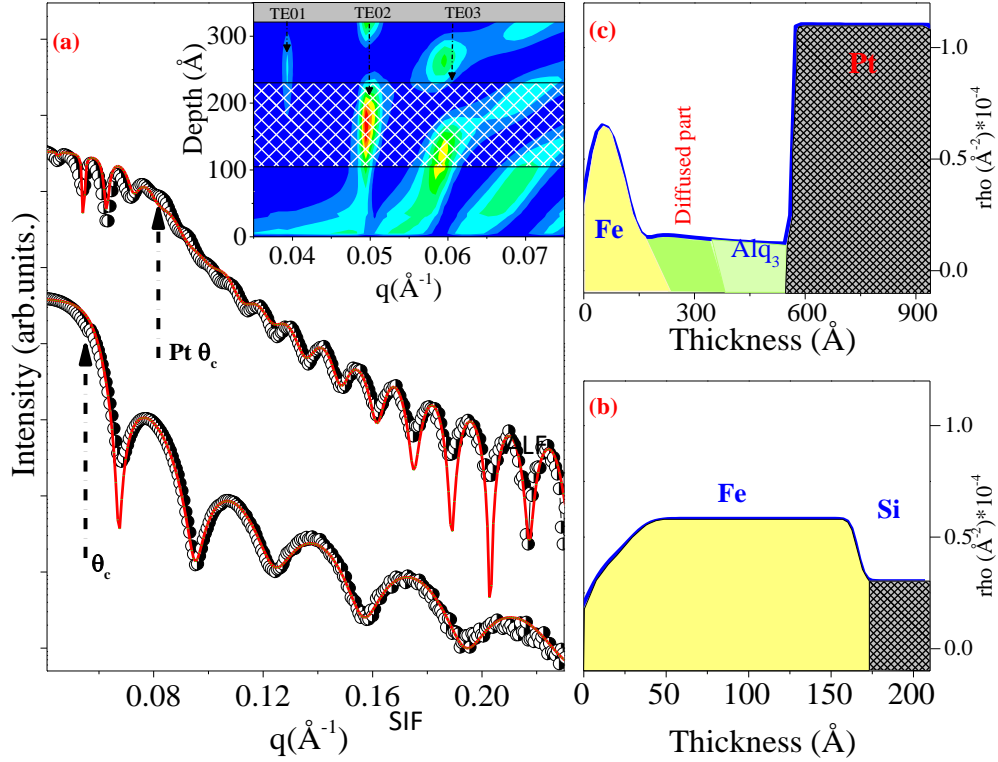

***Figure1***: *(a)Represents the x-ray reflectivity of Si/Fe (SIF) and Si/Pt/Alq3/Fe (ALF) as a function of scattering vector q. The continuous curves represent the best fit for the data. The inset of the figure shows simulated 2D x-ray intensity distribution during the formation of XSW as function of the x-ray penetration depth and the incident angle. Figures (b) and (c) represent the electron density depth profile for both samples, respectively.*

Thicknesses of Pt, $Alq_3$ layers in sample ALF are found as 405 Å and 290 Å, respectively. High-frequency Keising oscillations in the XRR pattern of the ALF sample as compared to the SIF sample are due to the higher total layer thickness (Pt+$Alq_3$+$^{57}$Fe+Fe) of the ALF sample. Drastic differences in both XRR patterns can be seen near the critical angle, where additional oscillation in the pattern is observed below $q_x$ ~0.083 Å$^{-1}$. These oscillations correspond to the excitation of different standing wave modes (TE0, TE1, TE2 etc.) in the $Alq_3$ layer[28,41]. These standing wave modes are always present in a low, dense layer when it is sandwiched between two high dense layers and known as waveguide structure. In the present case, such a condition is intentionally created by depositing Pt layer below $Alq_3$/$^{57}$Fe/Fe to confined XSW antinode near the diffused $Alq_3$/$^{57}$Fe interface and getting enhanced NRS yield from the diffused layer, which is discussed later. The inset of fig1(a) gives simulated x-ray intensity distribution inside the $Alq_3$ layer for the ALF sample, where the confinement of the x-ray intensity in different XSW modes are visual. This simulation is based on Parratt's formulism[25]. The position of the $^{57}$Fe/Fe layer is marked with respect to XSW modes. At about q=0.05Å$^{-1}$, the $Alq_3$/$^{57}$Fe interface overlaps with TE2 antinode, which provides the feasibility for enhanced interface contribution at this q values. In view of this fact, all GI-NRS measurements are performed (discussed later on) by keeping the incident angle at q=0.05 Å$^{-1}$.

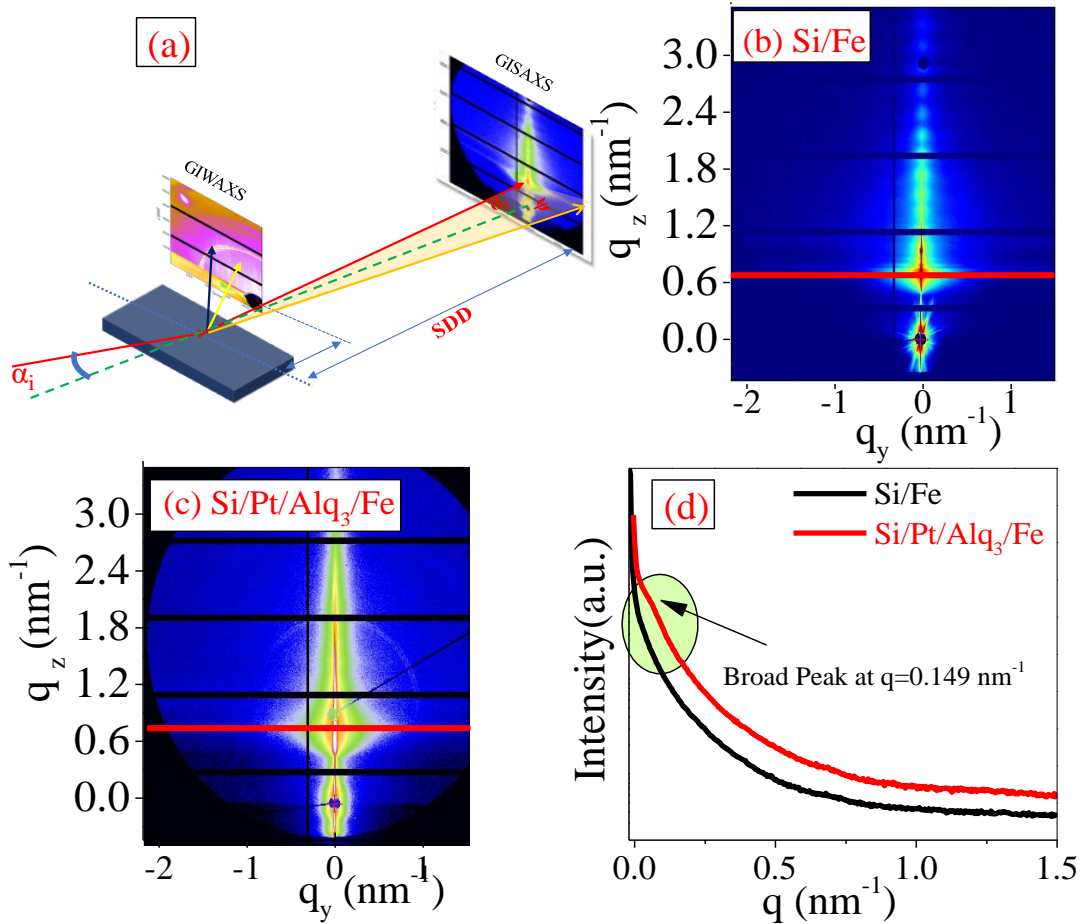

*Figure 2: (a) Shows the Schematic of geometry used for GISAXS and GIWAXS. 2D GISAXS data for (b) Si/57Fe/Fe (SIF) and (c) Si/Pt/Alq3/57Fe/Fe (ALF) samples (d) Integrated intensity profiles along qy direction for SIF and ALF samples.*

Further, GISAXS and GIWAXS measurements were performed simultaneously to study the morphology and structural growth of Fe on the $Alq_3$ layer. Photon energy of 13 keV was used to illuminate the sample at a grazing-incident angle, larger than the critical angle for total external reflection, at which X-rays penetrate the sample. The experimental geometry is as shown in fig 2(a). Figure 2(b) and (c) shows a typical 2D GISAXS pattern of SIF and ALF samples. The in-plane vector component $q_y$ represents the in-plane structure. As marked in fig 2 (b) and 2 (c), a horizontal stripe was extracted to get GISAXS scattered intensity Vs $q_y$ (nm$^{-1}$) using program DPDAK[42] and plotted in Fig 2(d). A weak broad peak at q= 0.1496 nm$^{-1}$ for ALF sample correspondence to the average interparticle distance of about 42 nm. Broadening in the peak is related to the formation of the finite grain size with broad particle size distribution[43] inside the $Alq_3$ layer near the Fe/$Alq_3$ interface. It is mainly due to the penetration of metal atoms in the mechanically soft $Alq_3$ layer.

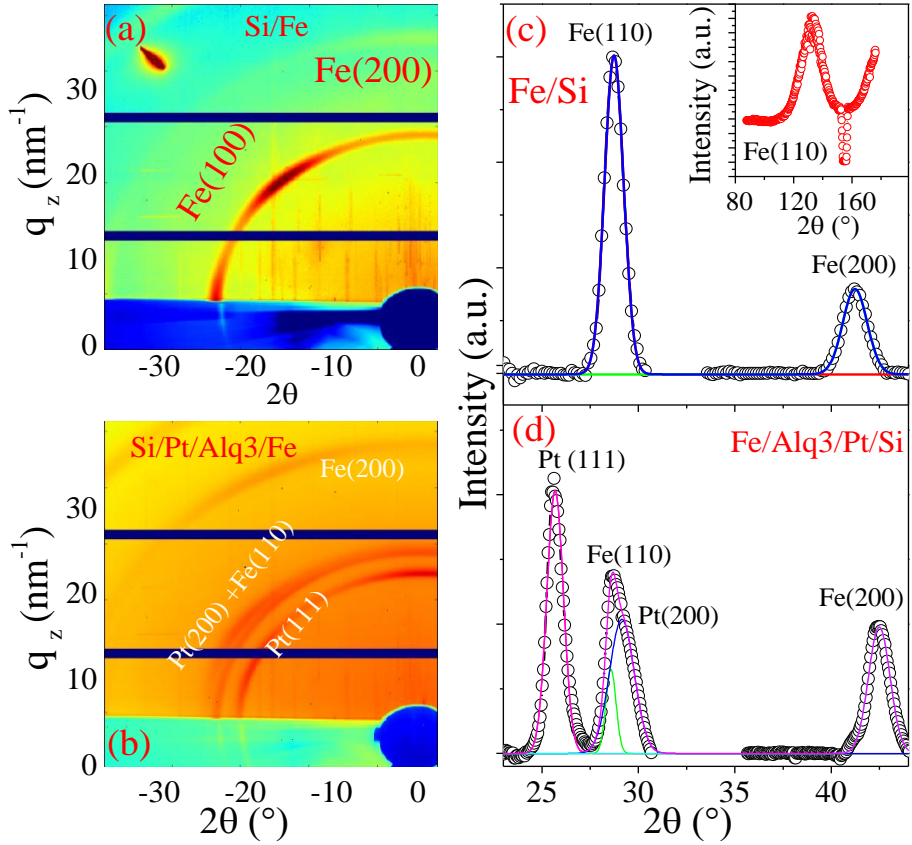

*Figure 3: (a) and (b) shows the 2D GIWAXS image for SIF and ALF, respectively. The extracted peak intensity profile integrated along the radial direction of the raw 2D GIWAXS patterns for (c) SIF and (d) ALF sample. Inset to (c) shows the azimuthal extraction for Fe (110) for SIF.*

After the GISAXS measurements, GI-WAXS measurements were performed in the same setting. For this purpose, as shown in fig 3 (a), another 2D detector (Pilatus 300K) along with GI-SAXS 2D detector was placed near the sample to cover a wide angler range ~ 90° to 180° to record GIWAXS data. Figure 3 (a) and (b) gives GIWAXS 2D diffraction images, collected for both the samples simultaneously with GISAXS by keeping X-ray beam at grazing incident angle ~0.4°. It may be noted that the bottom parts of different rings are shadowed by the substrate, whereas the right half part of the rings could not be collected due to the geometrical constraints in simultaneous GISAXS measurements. Nevertheless, the visual part of the rings is sufficient for our intended analysis of the sample structure. In the case of the SIF sample, discontinuous diffraction rings corresponding to (110) and (200) planes of bcc Fe layer confirm the textured Fe grains[44, 45]. An intense spot at a higher angle is due to the crystalline silicon substrate. Along with sharpening of the rings, Fe rings are found relatively continuous which confirms the presence of big particle size and weak texturing in Fe film on the $Alq_3$ layer (sample-ALF). Additional Pt ring in ALF sample is due to the buffer Pt layer, used to make waveguide structure for XSW based measurements. Overall integrated intensities of the concentric rings from both images, corresponding to diffraction planes (110) and (200) of peaks of Fe layer, are obtained by extracting range 27 to 40 degree using DPDAK[39] software and plotted in fig3 (c & d) In case of the ALF sample asymmetric peak around 27° is due to the overlapping of the Fe(110) and Pt (200). In order to separate Fe information, this broad peak was fitted with two peaks at positions 28.57° (Fe) and 29.14° (Pt).

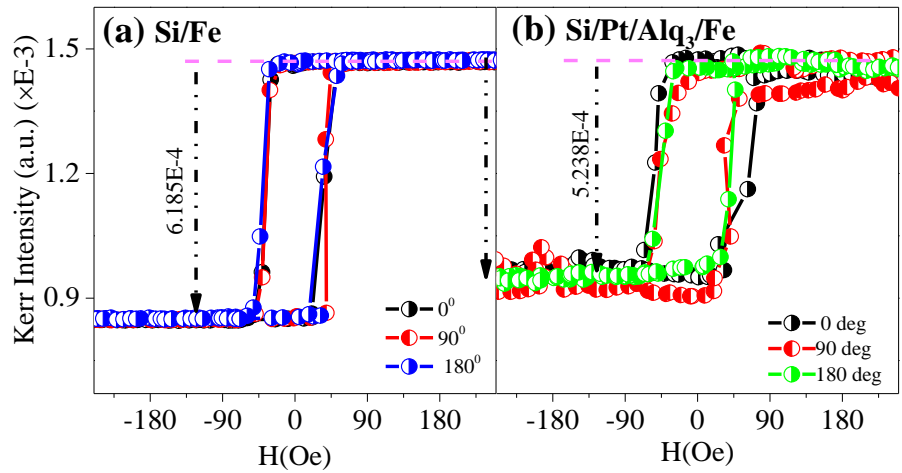

*Figure 4: Represents the MOKE hysteresis loop of (a) Si/Fe and (b) Si/Pt/Alq3/Fe structure.*

Fig 4 (a) and (b) shows the hysteresis loops of (a) SIF and (b) ALF obtained at an azimuthal angle of θ= 0°, 90° and 180° using MOKE measurement. Similar loops for the sample confirm the absence of any magnetic anisotropy in either case. On the other hand, in the case of the ALF sample, one may note that the Kerr signal (height of the loop) reduces about 15 % with an increase in coercivity of about 15%). This reduction in signal signifies that the magnetic contribution is reduced, whereas the increase in the coercivity is mainly due to the increase in pinning centers in the ALF sample. Mixing and diffusion at the Interface could be the main factor and responsible for such magnetization reversal with reduced Kerr signal. However, MOKE is incapable of resolving interface magnetism separately. To further investigate interface magnetism in detail, the same samples were prepared in identical conditions by placing a thin $^{57}$Fe layer at Fe/Alq$_3$ and Fe/Si interfaces and GINRS measurements were performed at the P01 beamline at PETRA III, Germany[39]. Being highly sensitive and an isotope sensitive technique, GINRS probes the $^{57}$Fe layer and provides magnetism from the $^{57}$Fe-Alq$_3$ and $^{57}$Fe-Si interfaces[28].

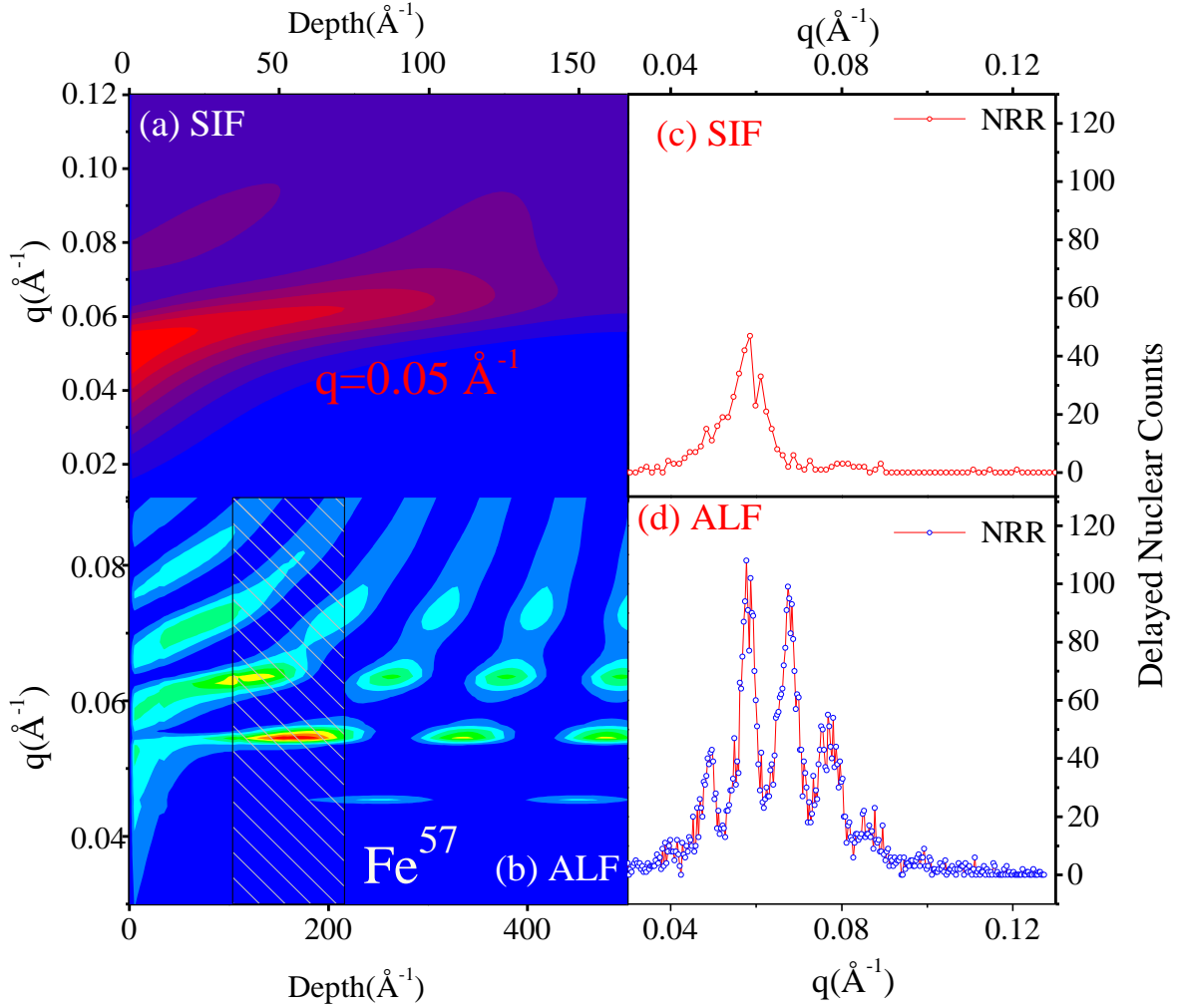

**Figure 5**: shows the electric field intensity distribution for (a) SIF and (b) ALF samples, respectively and (c) and (d) depicts the simulated NRR for SIF and ALF samples respectively.

Since the marker layer is very thin, therefore to select an appropriate incident angle to enhance resonant counts from the $^{57}$Fe marker layer, nuclear resonance counts have been collected with increasing incident angles for both samples. The selection of incident angle is confirmed using simulated 2D x-ray field intensity confinement region using parrot formulism, as shown in fig. 5 (a) and (b). Here, by selecting different incident angle, provides relative information from the different region of the film structure[28]. Figures 5 (c) and (d) are nuclear-delayed count Vs incident angle for both samples. In the case of the SIF sample, a maximum resonance of about 40 counts/sec is achieved at angle 0.05 Å$^{-1}$, Whereas, for the ALF sample, resonance enhancement is seen at various angles due to the formation of standing wave [28, 46]. Among all, a maximum resonant of about 110 counts /sec is achieved at an incident angle of 0.05 Å$^{-1}$. Interestingly, the formation of standing wave, about 3-fold enhancement of signal compared to the SIF sample is obtained for the ALF sample. GI-NRS measurements are performed for both samples by taking incident angle $q_{in}$=0.05 Å$^{-1}$ and presented in figures 6 (a) and (b). Oscillations (quantum beats) in resonant count in time spectra are due to the interference of electromagnetic waves that are emitted by different hyperfine components. These beating patterns in both samples are fitted using REFTIM software[47] and the magnetic information obtained from the $^{57}$Fe layer; hence information from the interfaces of Fe/$^{57}$Fe/Si and Fe/$^{57}$Fe/Alq$_3$ samples. To fit GINRS data of the ALF sample, the $^{57}$Fe layer at the Interface is divided into three sub-layers with different nuclear concentrations, hyperfine fields, and magnetic spin alignment. Hyperfine field of $^{57}$Fe layer along the depth of the Interface is found with reduced nuclear density and hyperfine fields (32T, 28T, and 12T), as compared to the hyperfine field value of ~32 T for pure bulk Fe. In the SIF sample, the hyperfine field of 32 T is recorded with a 25 T field that corresponds to the interface region. The resultant magnetic moment in SIF orients themselves out of plane 36° at the Interface with respect to the film plane. on the other hand, all moments orient themselves in the film plane in the case of ALF sample. Present results provide evidence for deep penetration of Fe atoms into Alq$_3$ film with the reduced hyperfine field at the buried region. The hyperfine field contribution is shown in pie diagram fig 6 (e) and (f). The buried part with field 28T contributes to 64% of the overall nuclear density contribution. 13% of $^{57}$Fe atoms show the nonmagnetic characteristic having 12T hyperfine field, while the presence of 32T field contributes to 23%. The total $^{57}$Fe HFI depth-profiles turn out to be in a good agreement with the measured XRR and MOKE measurements, where 15 % reduction of MOKE signal in the ALF sample is in agreement with GINRS measurements where 13% of the nonmagnetic contribution is found in the diffused Fe layer at the Interface. It is clear now that the reduced hyperfine field in the diffused part is due to the change in the chemical environment near the Fe or possible reduced cluster size at the buried part.

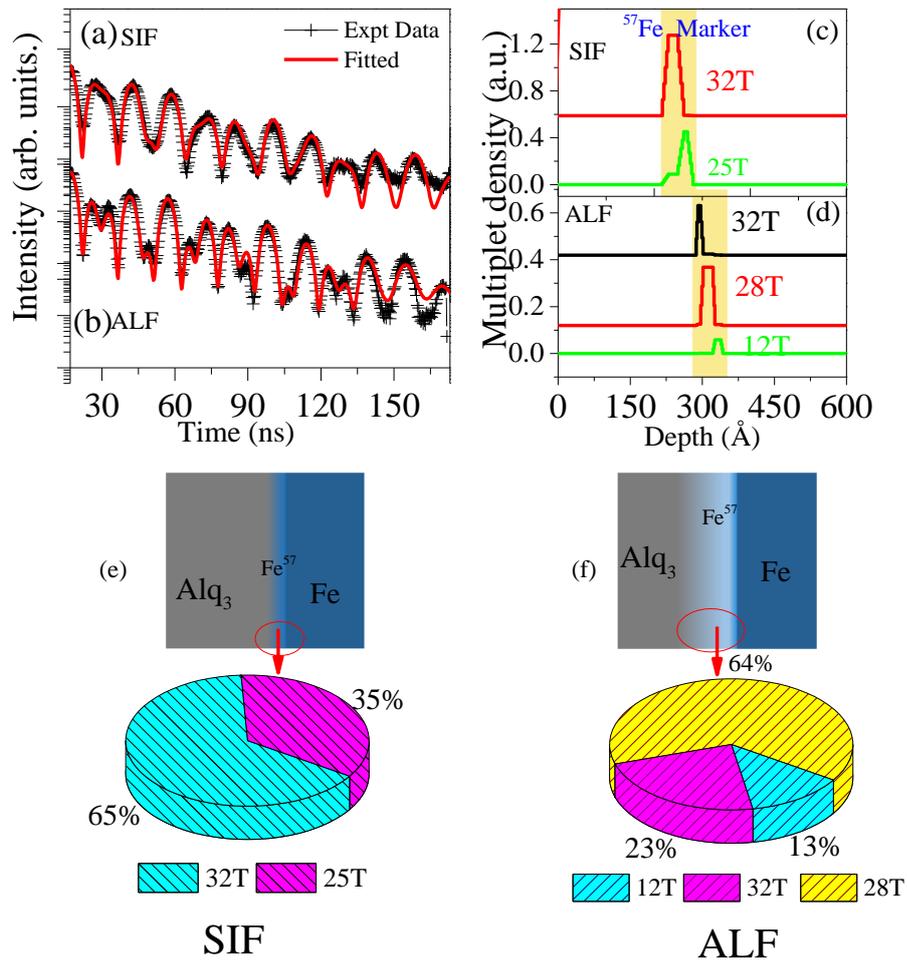

*Figure 6: GINRS time spectra for (a)SIF and (b) ALF systems. (c) and (d) Represents the Depth distribution of hyperfine multiple for SIF and ALF samples, respectively. The distribution of contribution of Hyperfine field in both SIF and ALF is shown in (e) and (f), respectively.*

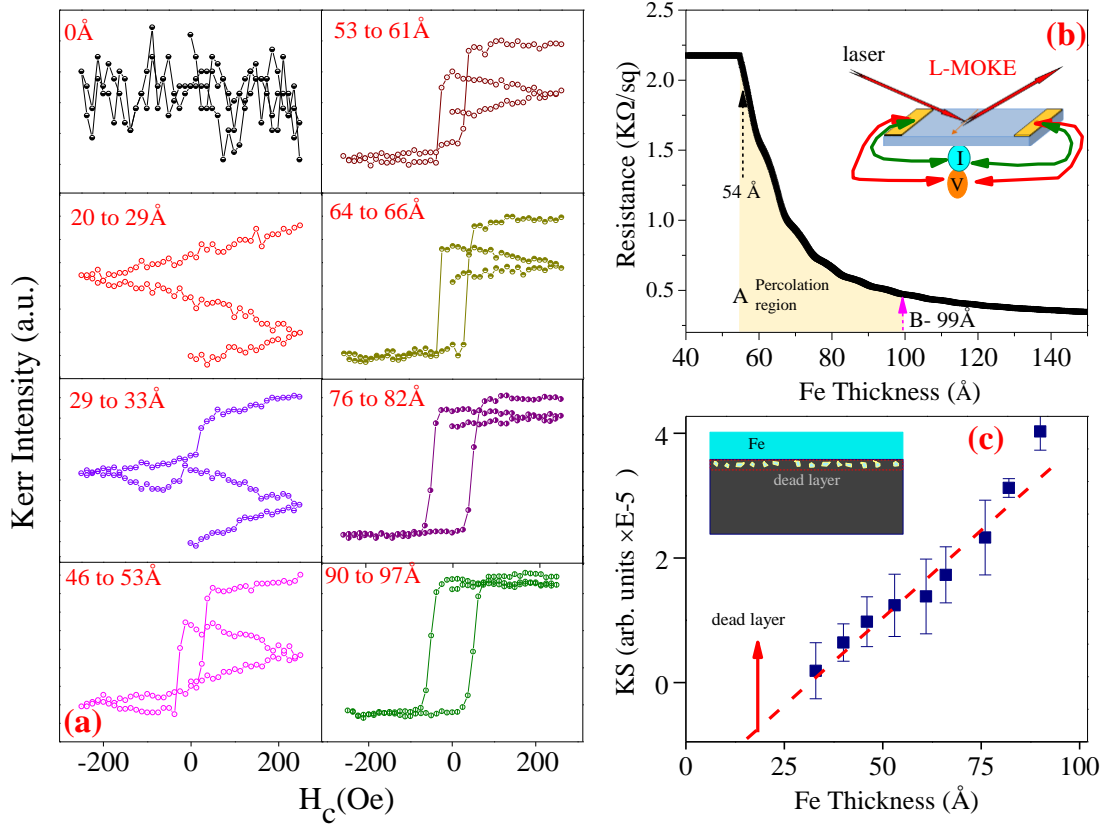

***Figure 7***: *(a) Some representative MOKE hysteresis loop showing the evolution of MOKE loop with thickness. (b): Sheet resistance RFe of Fe film Vs Fe film thickness. The percolation is started at 54.63 Å. Incent to fig (b) shows the geometry for measurement (c): Represents the variation of Kerr Signal with Fe thickness. indicating the MDL of 24 Å.*

To further understand the interface magnetism and to correlate it with the intermixing at the Interface of Fe-Alq$_3$, Fe films, starting from a fraction of nanometer to tens of nanometer, are grown on Alq$_3$ film in identical conditions and characterized simultaneously using in-situ MOKE and transport measurements. Some representative MOKE hysteresis loops collected at the different thicknesses and simultaneously obtained thickness-dependent sheet resistance are given in figures 7(a) and 7(b), respectively. It may be noted that the Kerr signal of the first reversal cycle of the loop is different from the second reversal cycle of the loop. This is mainly due to the MOKE measurements during deposition. It may be noted that it takes around 30 seconds to collect an entire hysteresis loop. Therefore, about 4 to 5 Å Fe thickness gets deposited on Alq$_3$[48] during this time. Whereas, at the higher thickness around 90 to 97 Å, the Kerr signal is less sensitive to the thickness due to limited penetration of the MOKE laser (He-Ne laser) and almost similar height of the loops are observed. Fig. 7 (b) gives the thickness dependence of sheet resistance (R) performed simultaneously with MOKE measurements. These measurements revealed that the film grows via Volmer–Weber mechanism, where islands grow larger to impinge with other islands. Eventually, coalescence starts around 5.5nm and shows a drastic reduction in sheet resistance[25]. Around 10 nm thickness, the slow decrease in resistance with thickness suggests the formation of a continuous layer[49]. Kerr signal extracted from the hysteresis loops was plotted as a function of thickness and shown in fig 7 (c). As expected, the variation of Kerr intensity is linear with film thickness. Extrapolation of the linear fit to the data is used to estimate the thickness of the magnetic dead layer at the Interface. In the present case, it is found to be 24 Å, depicting the nonmagnetic contribution at the Interface [50, 51, 52, 53]. The formation of the magnetically dead layer at the Interface could be related to the diffusion at the Interface and or the formation of paramagnetic nanoclusters.

## 4. CONCLUSIONS

A detailed study of Interface resolved magnetism and the topological structure of Fe on organic semiconductor Tris(8-hydroxyquinolinato) aluminium ($Alq_3$) is performed using GINRS under x-ray standing wave condition, GISAXS, GIWAXS and in-situ MOKE measurements. Compared with Fe growth on Si substrate, where film exhibits a hyperfine field value of ~32 T (Bulk Fe), a thick Fe-$Alq_3$ interface has been found with reduced electron density and hyperfine fields, providing evidence of deep penetration of Fe atoms into $Alq_3$ film. Due to soft nature of $Alq_3$, Fe moments relax themselves in the film plane, while Fe on Si has a resultant ~36° out of plane orientation of Fe moments at the Interface due to stressed and rough Fe layer near Si. In-situ MOKE during growth of Fe on $Alq_3$ surface and post-deposition GISAXS measurements revealed that Fe atoms try to organize into clusters to minimize their surface/interface energy. The origin of the 24 Å thick magnetic dead layer at the Interface is attributed to the small Fe clusters, which are of paramagnetic or superparamagnetic nature. Present work provides an understanding of interfacial magnetism at metal-organic interfaces and topological study using GI-NRS technique, which is made depth selective to probe magnetism of the diffused ferromagnetic layer effectively for special attention to the metal-organic interface frame, which otherwise difficult for lab-based techniques.

## ACKNOWLEDGEMENTS

Portions of this research were carried out at the light source, PETRA III of DESY, a member of Helmholtz Association (HGF). Financial support from the Department of Science and Technology (Government of India) (Proposal No. I-20180885) provided within the framework of the India@DESY collaboration is gratefully acknowledged. Mr. Shahid Jamal and Ms. Sadhana Singh are acknowledged for the help in In-situ measurements and discussion.